\documentclass[aps,pra,twocolumn,notitlepage,longbibliography,nofootinbib]{revtex4-1}

\usepackage{graphicx}
\usepackage{amsmath,amssymb,amsthm,amsfonts}
\usepackage{algorithm}
\usepackage{algorithmic}

\theoremstyle{plain}

\theoremstyle{definition}

\def\bF{\mathbb{F}}

\def\m{\mathop{\rm m}}
\def\ideal{\mathop{\rm ideal}}
\def\real{\mathop{\rm real}}
\def\mid{\mathop{\rm mid}}
\def\Pr{\mathop{\rm Pr}\nolimits}

\def\Label#1{\label{#1}\ [\ \text{#1}\ ]\ }
\def\Label{\label}

\begin{document}
\title{Optimum ratio between two bases in Bennett-Brassard 1984 protocol with second order analysis}
\author{Masahito Hayashi}
\email{hayashi@sustech.edu.cn}
\affiliation{Shenzhen Institute for Quantum Science and Engineering, Southern University of Science and Technology, Nanshan District,
Shenzhen 518055, China}
\affiliation{International Quantum Academy (SIQA), Futian District, Shenzhen 518048, China}
\affiliation{Guangdong Provincial Key Laboratory of
Quantum Science and Engineering, Southern University of Science and Technology, Nanshan District, Shenzhen 518055, China}
\affiliation{Graduate School of Mathematics, Nagoya University, Furocho, Chikusa-ku, Nagoya 464-8602, Japan}

\date{\today}

\begin{abstract}
Bennet-Brassard 1984 (BB84) protocol,
we optimize the ratio of the choice of two bases, the bit basis and the phase basis by using the second order expansion for the length of the generation keys
under the coherent attack.
This optimization addresses the trade-off between 
the loss of transmitted bits due to the disagreement of their bases
and the estimation error of the error rate in the phase basis.
Then, we derive the optimum ratio and the optimum length of the generation keys with the second order asymptotics.
Surprisingly, the second order has the order $n^{\frac{3}{4}}$, which is much larger than the second order $n^{\frac{1}{2}}$ in the conventional setting
when $n$ is the number of quantum communication.
This fact shows that our setting has much larger importance for the second order analysis than the conventional problem.
To illustrate this importance, 
we numerically plot the effect of the second order correction.
 \end{abstract}

\maketitle

\section{Introduction}
Bennet-Brassard 1984 (BB84) protocol \cite{BB84} is a standard protocol for quantum key distribution.
The key point of this protocol is 
the evaluation of the amount of information leakage on the bit basis via
the estimation of the error rate in the phase basis.
Due to this reason, the sender, Alice, and the receiver, Bob, choose 
their basis independently with equal probability in the conventional setting.
In this method, a half of the transmitted bits are discarded due to the disagreement of their bases.
However, since the aim is the estimation for the error rate,
it is sufficient to assign 
the phase basis to a limited number of transmitted pulses
that enables Alice and Bob to estimate the error rate in the phase basis \cite{LCA}.
In this situation, we need to address the trade-off between 
the loss of transmitted bits due to the disagreement of their bases
and the estimation error of the error rate in the phase basis.
To address this problem, 
we need to clarify the effect of the estimation error to the key generation rate.
The existing study \cite{Ha09} treated the estimation error in the large deviation framework.
While the large deviation method addresses the speed of convergence of the amount of information leakage, it cannot directly address the fix amount of information leakage.
Due to this reason, people in the community of quantum information
are interested in the latter formulation than the large deviation theory.
Fortunately, the existing studies \cite{Ha06,HT12}
investigated this trade-off problem in the security proof under the coherent attack 
by using the second order analysis
while the preceding studies \cite{SP,Mayers,Hamada,Renner,WMU} addresses only the first order analysis
in the asymptotic regime for the security proofs.
These studies \cite{Ha06,HT12} clarified that the order of the second order 
in the length of the key generation is $n^{\frac{1}{2}}$
when $n$ expresses the number of quantum communications.

The second order theory was initiated by Strassen \cite{Strassen}, and address the fixed amount of the error probability.
Then, the paper \cite{Ha06} applied it to the asymptotic regime of the security proof of QKD and, 
the paper \cite{Ha08-1} did it to the classical source coding and uniform random number generation.
However, this approach did not attract attention sufficiently
until the papers \cite{Ha09-8,PVV} applied it to the classical channel coding.
After the papers \cite{Ha09-8,PVV}, the papers \cite{ToH13,Li} applied this approach to other topics in quantum information.
In particular, the paper \cite{ToH13} studied the secure random number extraction and the data compression with quantum side information in this framework.
While the paper \cite{TLGR} studied the finite-length regime for the security proofs,
the paper \cite{HT12} established the bride between 
the finite-length and second order regimes for the security proofs.
That is, it derived the finite-length bound for key generation and 
recovered the second order asymptotics as its limit.
Later, the papers \cite{KMFBB,KKGW} considered the second order analysis for QKD under the collective attack, but they 
assumed that the error of the channel estimation is zero.
Overall, the order of the second order is 
$n^{\frac{1}{2}}$ when $n$ is the order of the first order.

In this paper, using the second order analysis under the coherent attack
by \cite{Ha06,HT12}, 
we address the trade-off between the loss of transmitted bits due to the disagreement of Alice's and Bob's bases
and the estimation error of the error rate in the phase basis.
Then, we optimize the ratio of the phase basis dependently of 
the observed error rates.
As the result, we find that the order of the second order in the length of the key generation is $n^{\frac{3}{4}}$ while
$n$ expresses the number of quantum communications.
Comparing the above existing studies, 
no preceding study derived the order $n^{\frac{3}{4}}$ as the second order.
Further, 
our second order $n^{\frac{3}{4}}$ is much larger than the conventional second order.
This fact shows that 
our problem has a larger effect by the second order correction, i.e.,
the second order analysis in our setting is more important than 
the second order analysis in other problem settings.
To clarify this importance, 
we numerically plot the effect of the second order correction.

The remaining part of this paper is organized as follows.
Section \ref{S2} states the optimum key generation length and 
 makes its numerical plot.
Section \ref{S3} shows the concrete protocol for our analysis
by combining the error verification.
Section \ref{S4} gives the detail derivation for our obtained result.

\section{Main results}\Label{S2}
In BB84 protocol, for each transmission, the sender, Alice,
randomly chooses one of two bases, the bit basis $\{|0\rangle, |1\rangle\}$ 
and the phase basis $\{|+\rangle, |-\rangle\}$, where
$|\pm\rangle := \frac{1}{\sqrt{2}}(|0\rangle \pm |1\rangle)$. 
The receiver, Bob, measures each received state
by choosing one of these two bases.
While these choices are done with equal probability in the usual case,
we assume that Alice and Bob choose the bit basis with probability $1-r_0 $.
Also, we assume that Alice and Bob choose the bit basis with probability $1-r_0$.
After their quantum communication, Alice and Bob find which 
quantum transmission is done in the matched basis by exchanging  
their basis choice via public communication.
While they keep the data in the matched basis,
they exchange a part of them to estimate the error rate.
Here, we denote the ratio of data used for estimation in 
the bit basis (the phase basis) by $r_1$ ($r_2$).

When the quantum channel is noisy,
we need information reconciliation and privacy amplification
after quantum communication.
Privacy amplification can be done by applying 
a typical type of hash function with calculation complexity $O(n\log n)$ where $n$ is the block length.  
Hence, 
we can choose the hash function dependently of 
the error rate of the channel.
In contrast, for a practical setting for BB84 protocol, we often fix our code
with coding rate $\beta$ for information reconciliation because it is not so easy to construct an error correcting code
dependently of the error rate of the channel.
In this paper, we adopt the following security criterion.
We denote Alice's and Bob's final keys by $K$ and $\hat{K}$, respectively, and denote 
Eve's system by $E$.
Also, we denote the public information and the length of final keys by $G$ and $L$.
In this situation, the ideal state $\rho^{ideal}_{LG K\hat{K}  E}$
is given by using $\vec{\sigma}_{E|LG}=(\sigma_{E|L=l,G=g})_{l,g}$
as follows.
\begin{align}
&\rho^{\ideal}_{LG K\hat{K} E}(\vec{\sigma}_{E|LG})\nonumber \\
:=& \sum_{l=0}^{l_{\m}}
\sum_{g} P_{LG}(l,g)|l,g \rangle \langle l,g|
\otimes 
\sum_{k=1}^{2^l} \frac{1}{2^l}|k,k\rangle \langle k,k| \nonumber \\
&\otimes 
\sigma_{E|L=l,G=g},
\end{align}
where $l_{\m}$ expresses the maximum length of final keys.
Therefore, our security criterion for our final state $\rho^{\real}_{LG K\hat{K} E}$ 
is given as the difference between the ideal state 
$\rho^{\ideal}_{LG K\hat{K} E}$ and the real state $\rho^{\real}_{LG K\hat{K} E}$
as 
\begin{align}
{\cal C}(\rho^{\real}_{LG K\hat{K} E})
:=\min_{\vec{\sigma}_E}
\frac{1}{2}\|\rho^{\ideal}_{LG K\hat{K} E}(\vec{\sigma}_{E|LG})-
\rho^{\real}_{LG K\hat{K} E}\|_1.
\end{align}
If $\vec{\sigma}_E$ is fixed to the state
$\vec{\rho}_{E|LG}=(\rho_{E|L=l,G=g})_{l,g}$, the above value is the same
as the criterion defined in \cite{Ben-Or}.
When we attach the error verification step, 
we can guarantee the correctness of our final keys
without caring about the estimation error of the error rate of the channel \cite[Section VIII]{Fung}.

We denote the final states for the part generated by the bit basis (the phase basis)
by $\rho^{\real,1}_{LG K\hat{K} E}$ ($\rho^{\real,2}_{LG K\hat{K} E}$).
Now, we impose our protocol to the condition under the coherent attack.
\begin{align}
{\cal C}(\rho^{\real,1}_{LG K\hat{K} E})\le \epsilon+o(\frac{1}{\sqrt{n}}), \quad
{\cal C}(\rho^{\real,2}_{LG K\hat{K} E})\le \epsilon+o(\frac{1}{\sqrt{n}}).
\Label{NNL}
\end{align}

Now, we employ the second order asymptotics for the generated key length 
\cite[Sections II-B and III-B]{Ha06} and \cite[Eq. (53)]{HT12}. 
When the observed error rates  in the bit basis (the phase basis) 
is given as $p_1$ ($p_2$) and the error verification is passed,
the averaged length of generated keys can be approximated by
\begin{align}
&n\Big(A(p_1)(1-r_0)^2 (1-r_1)+
A(p_2) r_0^2 (1-r_2)\Big) \nonumber \\
&-\sqrt{n} \Big(
B(p_2,\epsilon)\nonumber \\
&\quad \cdot \sqrt{\frac{(1-r_0)^2(1-r_1) (    (1-r_0)^2(1-r_1)+ r_0^2 r_2)}
{r_0^2 r_2} } \nonumber \\
&+
B(p_1,\epsilon)\sqrt{\frac{r_0^2(1-r_2) ( r_0^2(1-r_2)+ (1-r_0)^2 r_1)}
{(1-r_0)^2 r_1} }
\Big)\nonumber \\
&+o(\sqrt{n}),\Label{CNO}
\end{align}
where
\begin{align}
A(p):= \beta-h(p), 
~B(p,\epsilon):=h'(p) \sqrt{p(1-p)} \Phi^{-1}(\epsilon^2)
\end{align}
and $\Phi(x):=\int_x^{\infty} \frac{1}{\sqrt{2\pi}} e^{-t^2/2}dt$.
Here, $h(p)$ expresses the binary entropy $- p \log p-(1-p)\log (1-p) $, and
$h'(p)$ expresses its derivative.

When $ h(p_2) \le h(p_1)  $, 
the optimal choice of $r_0,r_1,r_2$ are 
$\sqrt{
\frac{B(p_2,\epsilon)}{2A(p_2)}
} n^{-\frac{1}{4}}$,
$0$, $1$.
The maximum averaged length of generated keys is
\begin{align}
& n A(p_2)-n ^{\frac{3}{4}}2 \sqrt{2 A(p_2) B(p_2,\epsilon)}
+O(n^{\frac{1}{2}}) \nonumber \\
=&n A(p_2) \Big(1  - n ^{-\frac{1}{4}} 2 \sqrt{ \frac{2 B(p_2,\epsilon)}{A(p_2) }} +O(n^{-\frac{1}{2}})\Big).
\Label{N2}
\end{align}
After this optimization, the second order has the order $n^{\frac{3}{4}}$, 
which is a larger order than the second order in \eqref{CNO}. Fig.
1 shows the optimum key generation rate with the second order correction when $p_2 = 0.05$. 
Since the second order $n^{\frac{1}{4}}$ appears in the rate, its effect is not negligible up to $n = 10^{10}$. 
This phenomena is surprising in comparison with the conventional second order analysis because the second order $n^{\frac{1}{2}}$ appears in the rate in the conventional setting
so that its effect vanishes around $n = 10^5$. 
This fact shows that the second order correction is more important when
we optimize the ratios $r_0, r_1, r_2$ in our 
modified BB84 protocol given as Protocol 1 than the conventional case.

\begin{figure}[t]
	\centering 
  \includegraphics[width=0.95\linewidth]{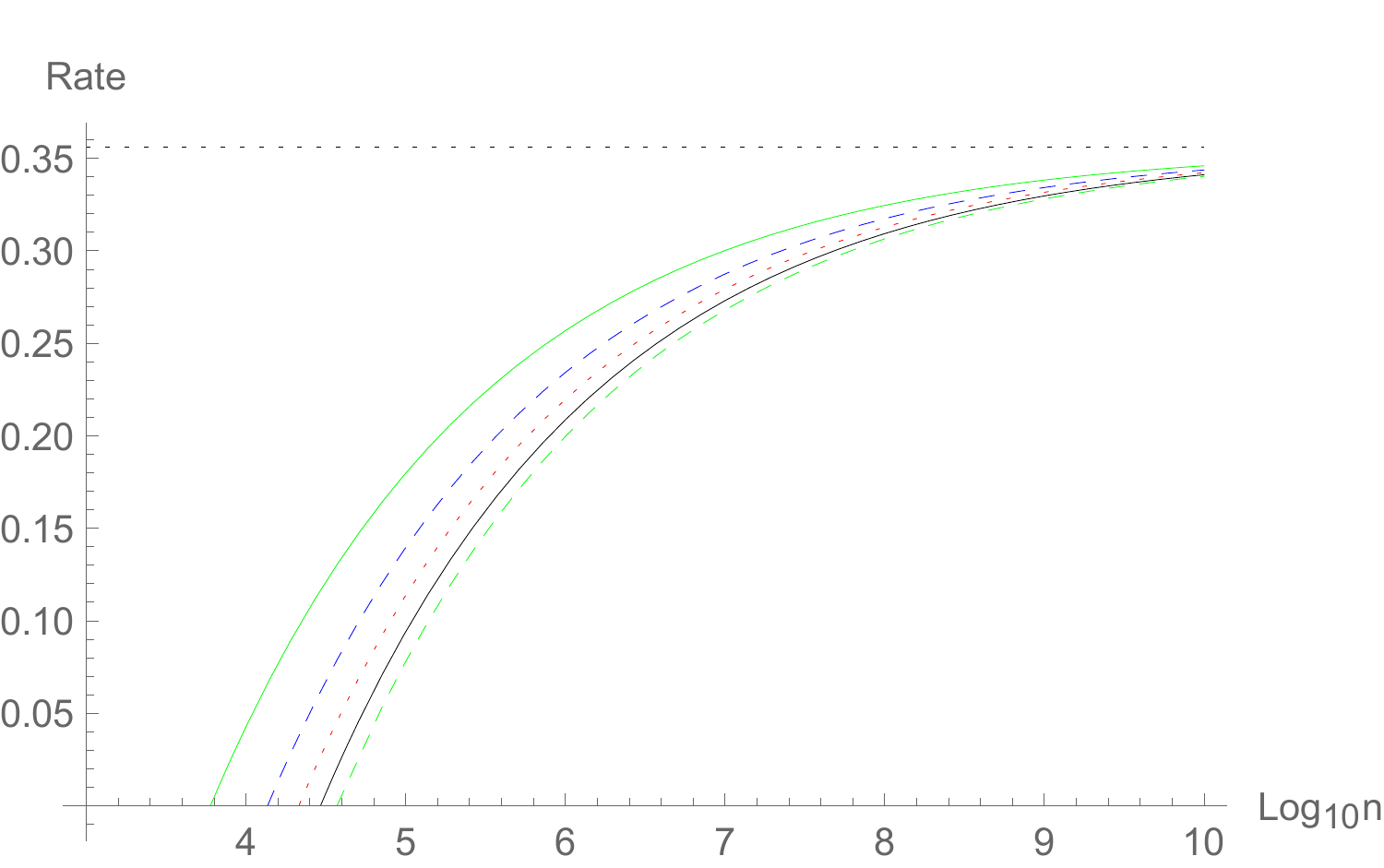}
	\caption{Numerical plot of the key generation rate 
	$A(p_2) \Big(1  - n ^{-\frac{1}{4}} 2 \sqrt{ \frac{2 B(p_2,\epsilon)}{A(p_2) }} \Big)$ with $p_2=0.05$
	and $\beta= 0.9 (1-h(0.05))=0.642243$.
The vertical axis expresses the rate, and the horizontal axis expresses the $\log_{10} n$. 
The top black dotted line expresses the first order rate, i.e.,  $A(0.05)=0.355846$.
The green normal line expresses the case with $\epsilon=10^{-2}$.
The blue dashed line expresses the case with $\epsilon=10^{-4}$.
The red dotted line expresses the case with $\epsilon=10^{-6}$.
The black normal line expresses the case with $\epsilon=10^{-8}$.
The green dashed line expresses the case with $\epsilon=10^{-10}$.
	\Label{fig}}
\end{figure}

\section{Detail description of our protocol}\Label{S3}
To show our main result,
we state our protocol.
This protocol uses modified Toeplitz matrices in privacy amplification.
A randomized function $f_S$ with random seeds $S$ is called 
a modified Toeplitz matrix from $\bF_2^{l_1}$ to $\bF_2^{l_2}$ with $l_1 \ge l_2$
when 
$S $ takes values in $\bF_2^{l_1-1}$ and 
$f_S$ is given as the matrix $(I, T(S))$, where
$T(S) $ is the $l_2 \times (l_1-l_2)$ Toeplitz matrix, whose components are defined as
$T(S)_{i,j}= S_{j-i+l_1}$. 
In fact, a modified Toeplitz matrix $f_S$ is an example of 
universal2 hash functions \cite[Appendix II]{Ha11}.
Here, a randomized function $f_S$ from ${\cal X}$ to ${\cal Y}$ with random seed $S$ is called 
a universal2 hash function when
the condition 
\begin{align}
\Pr (f_S(x)=f_S(x') )\le \frac{1}{|{\cal Y}|}\Label{uni2}
\end{align}
holds for any $x\neq x' \in {\cal X}$ \cite{CW}.

Also, based on \cite[Sections II-B and III-B]{Ha06} and \cite[Eq. (4)]{HT21},
we define the small value;
\begin{align}
\delta(p,\epsilon,m_1,m_2):=
\sqrt{\frac{p(1-p)(m_1+m_2) }{m_1 m_2}}
 \Phi^{-1}(\epsilon_{du})\Label{XL1}
 \end{align}
with $\epsilon =\sqrt{ \epsilon_{du}}$.
That is, $\delta(p,\epsilon,m_1,m_2)$ is given as
\begin{align}
\delta(p,\epsilon,m_1,m_2)=
\sqrt{\frac{p(1-p)(m_1+m_2) }{m_1 m_2}}
 \Phi^{-1}(\epsilon^2).
 \end{align}
Then, our protocol is given as Protocol 1.

{\bf Protocol 1}
\Label{protocol1-0}
\begin{algorithmic}
\STATE {\bf Quantum Communication:}\quad 
Alice randomly chooses the bit basis or the phase basis with the ratio $1-r_0:r_0$
and sends $n$ qubits and Bob measures the $n$ receiving qubits by choosing the bit basis or the phase basis with the ratio $1-r_0:r_0$. 
Here, Alice chooses her bits subject to the uniform distribution. 
After quantum communication, they exchange the choice of bases via public channel.  
Then, they obtain $N_1=n_1$ bits with the bit basis and $N_2=n_2$ bits with the phase basis.
\STATE {\bf Error estimation:}\quad 
They randomly choose check bits in the bit basis (the phase basis) with ratio $r_1$ ($r_2$),
and obtain the estimate $p_1$ ($p_2$) by exchanging their information.
Then, they decide the sacrificed lengths 
$m_{1}(n_1,p_2):=(1-r_1)n_1( h(p_2+\delta(p_2,\epsilon,(1-r_1) n_1,r_2 n_2)))$ 
 and $m_{2}(n_2,p_1):=(1-r_2)n_2( h(p_1+\delta(p_1,\epsilon,(1-r_2) n_2,r_1 n_1)))$.
\STATE {\bf Information reconciliation:}\quad 
They apply error correction with the linear code $C_1$ ($C_2$) of the rate $\beta$ in the remaining bits in 
the bit basis (the phase basis).
That is, Alice sends her syndrome of the linear code $C_1$ ($C_2$) of 
$(1-r_1)n_1$ bits with the bit basis ($ (1-r_2)n_2$ bits with the phase basis) to Bob via public channel.
Bob corrects his error.
Then, Alice (Bob) obtains $\beta (1-r_1)n_1$ bits $X_1$ ($\hat{X}_1$) with the bit basis and 
$\beta (1-r_2)n_2$ bits $X_2$ ($\hat{X}_2$) with the phase basis.
\STATE {\bf Privacy amplification:}\quad 
Alice randomly chooses two modified Toeplitz matrices $f_{1,S_1}$ 
from $\beta (1-r_1)n_1$ bits to $\beta (1-r_1)n_1-m_{1} $ bits 
and $f_{2,S_2}$ from $\beta (1-r_2)n_2$ bits to $\beta (1-r_2)n_1-m_{2} $ bits,
and sends the choices of $S_1$ and $S_2$ to Bob via public channel. 
Then, Alice (Bob) obtains $f_{1,S_1}(X_1)$ ($f_{1,S_1}(\hat{X}_1)$) with the bit basis 
and $f_{2,S_2}(X_2)$ ($f_{2,S_2}(\hat{X}_2)$) with the phase basis. 

\STATE {\bf Error verification:}\quad 
Alice sets $m_{3}$ to be $\log n$.
Alice randomly chooses two modified Toeplitz matrices 
$f_{3,S_3}$ 
from $\beta (1-r_1)n_1-m_{1} $ bits to $m_{3} $ bits
and $f_{4,S_4}$ 
from $\beta (1-r_2)n_2-m_{2} $ bits to $m_{3} $ bits,
and sends the choices of $S_3$, $S_4$, and 
$f_{3,S_3}(f_{1,S_1}(X_1))$, $f_{4,S_4}(f_{2,S_2}(X_2))$
to Bob via public channel. 
If the relation
$f_{3,S_3}(f_{1,S_1}(X_1))=f_{3,S_3}(f_{1,S_1}(\hat{X}_1))$
($f_{4,S_4}(f_{2,S_2}(X_2))=f_{4,S_4}(f_{2,S_2}(\hat{X}_2))$) holds,
they keep their bits $f_{1,S_1}(X_1)$ and $f_{1,S_1}(\hat{X}_1)$ 
($f_{2,S_2}(X_2)$ and $f_{2,S_2}(\hat{X}_2)$) by discarding 
initial $m_{3}$ bits of 
$f_{1,S_1}(X_1) $ and $f_{1,S_1}(\hat{X}_1)$
($f_{2,S_2}(X_2) $ and $f_{2,S_2}(\hat{X}_2)$).
Otherwise, they discard their obtained keys, i.e., set the length $L$ to be zero.
\end{algorithmic}

\section{Derivation of our evaluation}\Label{S4}
For our security analysis under the coherent attack, 
we define the state
\begin{align}
&\rho^{\mid,i}_{LG K\hat{K} E}\nonumber \\
:= &
\sum_{l=0}^{l_{\m}}
\sum_{g} P_{LG}^i(l,g)|l,g \rangle \langle l,g|\nonumber \\
&\otimes 
\sum_{k=1}^{2^l}\frac{1}{2^l} |k,k\rangle \langle k,k|
 \otimes \rho_{E| K=k,L=l,G=g}^i 
\end{align}
for $i=1,2$.
As explained in Appendix \ref{A3}, using the property \eqref{uni2}, we can show
\begin{align}
\frac{1}{2}\|\rho^{\mid,i}_{LG K\hat{K} E}-\rho^{\real,i}_{LG K\hat{K} E}\|_1
\le \frac{1}{2^{m_{3}}}
=\frac{1}{n}\Label{NA1}
\end{align}
for $i=1,2$. Thus,
we expand the security criterion ${\cal C}( \rho^{\real,i}_{LG K\hat{K} E})
$ as
\begin{align}
&{\cal C}( \rho^{\real,i}_{LG K\hat{K} E})\nonumber \\
\le &
\frac{1}{2}\|\rho^{\ideal,i}_{LG K\hat{K} E}(\vec{\sigma}_{E|LG})
-\rho^{\mid,i}_{LG K\hat{K} E}\|_1\nonumber \\
&+
\min_{\vec{\sigma}_{E|LG}}
\frac{1}{2}\|\rho^{\mid,i}_{LG K\hat{K} E}-\rho^{\real,i}_{LG K\hat{K} E}
\|_1 \nonumber \\
\le &
\min_{\vec{\sigma}_{E|LG}}
\frac{1}{2}\|\rho^{\ideal,i}_{LG K E}(\vec{\sigma}_{E|LG})-\rho^{\real,i}_{LG K E}
\|_1
+\frac{1}{n}.\Label{NL1}
\end{align}

The papers \cite{Ha06,Ha07,HT12} considered
the virtual decoding error probability in the dual basis, which is denoted by
$P_{du}^i$ for $i=1,2$.
As shown in Appendix \ref{A1}, we have
\begin{align}
& 
\min_{\vec{\sigma}_{E|LG}}
\frac{1}{2}\|\rho^{\ideal,i}_{L G K E}(\vec{\sigma}_{E|L G})
-\rho^{\real,i}_{LG K E}\|_1
\le \sqrt{P_{du}^i}.\Label{NL2}
\end{align}
Now, we recall the result for the second order analysis by 
\cite[Sections II-B and III-B]{Ha06} and \cite[Eq. (4)]{HT21},
which is the corrected version of \cite[Eq. (53)]{HT12}.
Due to the choices of $m_{1}$ and $m_{2}$,
the above mentioned second order analysis guarantee that 
\begin{align}
P_{du}^i\le \epsilon_{du}+o(\frac{1}{\sqrt{n}})\Label{NL3}
\end{align}
under the coherent attack.
Since $\epsilon^2=\epsilon_{du}$,
combining \eqref{NL1}, \eqref{NL2}, and \eqref{NL3}, we have
\begin{align}
{\cal C}( \rho^{\real,i}_{LG K\hat{K} E})
\le &
\epsilon+o(\frac{1}{\sqrt{n}}),\Label{NL4}
\end{align}
which guarantees \eqref{NNL}.
That is, we find that Protocol 1 satisfies the condition \eqref{NNL}.

As shown in Appendix \ref{A4},
by using the definition of $\delta(p,\epsilon,m_1,m_2)$ given in \eqref{XL1}
the length of the generated keys is calculated as
\begin{align}
&\beta (1-r_1)n_1-m_{1}(n_1,p_2)-m_{3} \nonumber \\
&+
\beta (1-r_2)n_2-m_{2}(n_2,p_1)-m_{3}\nonumber \\
=&(1-r_1)( \beta -h(p_2))n_1 + (1-r_2)( \beta -h(p_1))n_2\nonumber  \\
&-B(p_2,\epsilon) \sqrt{\frac{((1-r_1) n_1+r_2n_2)(1-r_1) n_1}{r_2 n_2}} \nonumber \\
&-B(p_1,\epsilon) \sqrt{\frac{((1-r_2) n_2+r_1n_1) (1-r_2) n_2}{r_1 n_1}} 
+o(\sqrt{n}).\Label{EA7}
\end{align}
Since $n_1$ and $n_2$ are the realizations of the random variables $N_1$ and $N_2$,
we consider the average with respect to these variables.
Since the averages of $N_1$ and $N_2$ are $n(1-r_0)^2$ and $n r_0^2$, we have
\begin{align}
&\mathbb{E}_{N_1,N_2} \big[\beta (1-r_1)N_1-m_{1}(N_1,p_2)-m_{3} \nonumber \\
&+\beta (1-r_2)N_2-m_{2}(N_2,p_1)-m_{3}\big]\nonumber \\
=&\mathbb{E}_{N_1,N_2} \Bigg[(1-r_1)( \beta -h(p_2))N_1 \!+\! (1-r_2)( \beta -h(p_1))N_2 \nonumber \\
&-(1-r_1) B(p_2,\epsilon) \sqrt{\frac{((1-r_1) N_1+r_2N_2)(1-r_1) N_1}{r_2 N_2}} 
\nonumber \\
&-(1-r_2) B(p_1,\epsilon) \sqrt{\frac{((1-r_2) N_2+r_1 N_1)(1-r_2) N_2}{r_1 N_1}}\Bigg] \nonumber \\
&+o(\sqrt{n}) \nonumber \\
=&(1-r_1)( \beta -h(p_2))(1-r_0)^2 n + (1-r_2)( \beta -h(p_1)) r_0^2 n  \nonumber \\
&\!-\! B(p_2,\epsilon) \sqrt{\frac{((1-r_1) (1-r_0)^2+r_2r_0^2)(1-r_1) (1-r_0)^2}{r_2 r_0^2}}\nonumber \\
&\cdot \sqrt{n}\nonumber \\
&\!-\!B(p_1,\epsilon) \sqrt{\frac{((1-r_2) r_0^2+r_1 (1-r_0)^2)(1-r_2) r_0^2}{r_1 (1-r_0)^2}}\sqrt{n} 
\nonumber \\
&+o(\sqrt{n}),
\end{align}
which implies \eqref{CNO}.

Next, we optimize 
the ratios $r_0,r_1,r_2$ under the condition
$ h(p_2) \le h(p_1)  $.
In this case, the optimal rate in the first order coefficient is 
$( \beta -h(p_2))$.
To achieve this rate, the ratio $r_0$ needs to approach to $0$.
We set $r_0$ to be $\alpha_1 n^{-\frac{1}{4}}$.
Then, the above value is calculated as
\begin{align}
&
(1-r_1)A(p_2) n 
- 2 (1-r_1) A(p_2)\alpha_1 n^{3/4} \nonumber \\
&- B(p_2,\epsilon) \sqrt{\frac{(1-r_1)^2}{r_2 \alpha_1^2}}  n^{3/4} 
+ O( \sqrt{n})  \nonumber \\
=&
(1-r_1) A(p_2) n \nonumber \\
&-\Big( 2 (1-r_1)A(p_2)\alpha_1  
+ B(p_2,\epsilon) \frac{(1-r_1)}{r_2^{1/2} \alpha_1} \Big) n^{3/4} 
\nonumber \\
&+ O( \sqrt{n}) .
\end{align}
To maximize the first order coefficient, $r_1$ needs to be $0$.
 The maximum of 
 $-\Big( 2 A(p_2)\alpha_1  
+ B(p_2,\epsilon) \frac{1}{r_2^{1/2} \alpha_1} \Big)$
is realized when $r_2=1$ and $\alpha_1=\sqrt{
\frac{B(p_2,\epsilon)}{2A(p_2)}
} $. Under this choice, the above value equals \eqref{N2}.

\section{Discussion and conclusion}
We have derived the optimum key generate rate
when we optimize the ratios of basis choices.
Then, we clarified the second order effect under this optimization.
 While the second order has the order $n^{-\frac{1}{2}}$ in the key generation rate 
under the conventional setting,
the second order has the order $n^{-\frac{1}{4}}$ in the key generation rate in our setting.
Since the vanishing speed of the second order effect is quite slow in our setting,
we need to be careful for the effect by the second order correction.
Overall, our result has clarified that
the order of the second order becomes large after the optimization for 
the ratio of the choices of the bases.
Further, we can expect similar phenomena in a problem with a certain optimization.
That is, this result suggests a possibility that an optimization makes the order of the second order larger than 
the original order of the second order.

Our model assumes a single-photon source.
Many reports for implementation of quantum key distribution used weak coherent sources.
Unfortunately, our result cannot be applied to such practical systems while
decoy BB84 methods and continuous variable method can be used for such practical systems \cite{Decoy1,Decoy2,Decoy3,Decoy4,CV1,CV2,CV3}.
For practical use, we need to expand our analysis to the above two methods.
In our result, one basis is used to generate the sifted keys
and the other basis is used to estimate the quantum channel.
This idea can be generalized to the following;
We optimize the ratio among the pulses to generate the sifted keys and the pulses 
to estimate the quantum channel.
Therefore, we need to apply the above optimization to the above practical settings.
It is an interesting future study to clarify the order of the second order larger 
after the above optimization in such practical settings.

Next, we discuss the implementation cost for our protocol in the software part.
The numerical plots in Fig. \ref{fig} shows that 
the block length $n$ needs to be chosen as $10^{10}$ to attain 
the rate $ A(p_2)$.
However, it does not require to prepare an error correcting code with such a long block length.
It is sufficient to prepare modified Toeplitz matrices with such a long block length. 
This construction can be done only with the calculation complexity $O(n \log n)$
The reference \cite[Appendices C and D]{HT16} explains how to implement the multiplication of 
Toeplitz matrix.
Indeed, the reference \cite[Appendix E-A]{HT16} reported its actual implementation for key length $10^8$
using a typical personal computer equipped with a 64-bit CPU (Intel Core i7) with
16 GByte memory, and using a publicly available software library.
Therefore, we can expect to implement the privacy amplification with $n=10^{10}$
in a current technology.

Here, we should remark the relation between our method for privacy amplification
and the method by \cite{Renner,TLGR,TH13}.
Our method is based on the method by \cite{Ha06,Ha07,HT12},
and the paper \cite{TH13} clarified what condition for hash functions 
is essential for this method.
To clarify the point,
the paper \cite{TH13} introduced the concept of dual universal2 hash functions,
and explained the difference between dual universal2 hash functions
and universal2 hash functions, which are used in 
the method by \cite{Renner,TLGR,TH13}.
While the privacy amplification in our method \cite{Ha06,Ha07,HT12} requires 
a surjectivity and linearity, 
the privacy amplification in \cite{Renner,TLGR,TH13} works with a general universal2 hash function, i.e.,
the linearity is not needed in \cite{Renner,TLGR,TH13}.
However, as explained in \cite[Section III-C]{HT16}, 
our method has a better robustness than 
the method by \cite{Renner,TLGR,TH13}.

\section*{Acknowledgments}
The author was supported in part by the National Natural Science Foundation of China (Grant No. 62171212) and
Guangdong Provincial Key Laboratory (Grant No. 2019B121203002).

\appendix

\section{Proof of \eqref{NA1}}\Label{A3}
The relation \eqref{NA1} is shown as follows.
\begin{widetext}
\begin{align}
&\frac{1}{2}\|\rho^{\mid,i}_{LG K\hat{K} E}-\rho^{\real,i}_{LG K\hat{K} E}\|_1\nonumber \\
=&
\frac{1}{2}\Big\|\sum_{l=0}^{l_{\m}}
\sum_{g} P_{LG}^i(l,g)|l,g \rangle \langle l,g|
\otimes 
\sum_{k=1}^{2^l}\frac{1}{2^l} 
\Big(|k,k\rangle \langle k,k |
-\sum_{\hat{k}=1}^{2^l} P_{\hat{K}|K,L=l}^i(\hat{k}|k) 
|k,\hat{k}\rangle \langle k,\hat{k}|\Big)
 \otimes \rho_{E| K=k,L=l,G=g}^i\Big\|_1 \nonumber \\
=&
\frac{1}{2}\sum_{l=0}^{l_{\m}}
P_L^i(l)
\Big\|
\sum_{k=1}^{2^l}\frac{1}{2^l} 
\Big(|k,k\rangle \langle k,k |
-\sum_{\hat{k}=1}^{2^l}P_{\hat{K}|K,L=l}^i(\hat{k}|k) 
|k,\hat{k}\rangle \langle k,\hat{k}|\Big)
\Big\|_1 \nonumber \\
=&
P_{K,\hat{K}}^i ( \hat{K}\neq K ) \nonumber \\
\le &
\Pr \big(
f_{i,S_i}(X_i)\neq f_{i,S_i}(\hat{X}_i) ,~
f_{2+i,S_{2+i}}(f_{i,S_i}(X_i))= f_{2+i,S_{2+i}}(f_{i,S_i}(\hat{X}_i)) 
 \big) \nonumber \\
= &
\Pr \big(
f_{i,S_i}(X_i)\neq f_{i,S_i}(\hat{X}_i)  \big) 
\Pr \big(
f_{2+i,S_{2+i}}(f_{i,S_i}(X_i))= f_{2+i,S_{2+i}}(f_{i,S_i}(\hat{X}_i))| f_{i,S_i}(X_i)\neq f_{i,S_i}(\hat{X}_i) 
 \big) \nonumber \\
\stackrel{(a)}{\le} &
\Pr \big(
f_{i,S_i}(X_i)\neq f_{i,S_i}(\hat{X}_i)  \big) 
\frac{1}{2^{m_{3}}}
 \le \frac{1}{2^{m_{3}}}
=\frac{1}{n},
\end{align}
where $(a)$ follows from \eqref{NA1}.
\end{widetext}

\section{Proof of \eqref{NL2}}\Label{A1}
To show \eqref{NL2}, 
we divide the public information $G$ into two parts $G_1$ and $G_2$.
$G_1$ is the public information except for 
$f_{2+i,S_{2+i}}(f_{2+i,S_{2+i}}(X_i))$
and 
$G_2$ is the public information $f_{2+i,S_{2+i}}(f_{2+i,S_{2+i}}(X_i))$.x
Also, we denote keys after Privacy amplification and its length by 
$K_*=(K_1,K_2)$ and $L_1$,
respectively, where
$K_1$ is the initial $m_{3}$ bits and $K_2$ is the remaining bits.
Since $K_1\mapsto f_{2+i}(K_1 k_2)$ is bijective for every $k_2$,
$(K_1,K_2)$ and $(G_2,K_2)$ have a one-to-one relation.
 Now, we say that the phase basis (the bit basis) is the dual basis
when we focus on the information on the bit basis (the phase basis).
That is, when $i=1$ ($i=2$), the dual basis is the phase basis (the bit basis). 

Now, we focus on the fidelity 
$F(\rho^{\ideal,i}_{L_1G_1 K_* E}(\vec{\sigma}_{E|L_1 G_1}),
\rho^{\real,i}_{L_1G_1 K_* E})$
between
$\rho^{\ideal,i}_{L_1G_1 K_* E}(\vec{\sigma}_{E|L_1 G_1})$ and 
$\rho^{\real,i}_{L_1G_1 K_* E}$.
We define the virtual decoding error probability $P_{du|L_1=l}^i$ 
 in the dual basis
 for $i=1,2$ dependently of $L_1=l$.
As shown in Appendix \ref{A2}, the relation
\begin{align}
&\max_{\vec{\sigma}_{E|G_1}}
F(\rho^{\ideal,i}_{G_1 K_* E|L_1=l}(\vec{\sigma}_{E| G_1}),
\rho^{\real,i}_{G_1 K_* E |L_1=l})\nonumber \\
\ge & \sqrt{1- P_{du|L_1=l}^i} \Label{NM1}
\end{align}
holds.
Hence, we have
\begin{align}
&\max_{\vec{\sigma}_{E|L_1 G_1}}
F(\rho^{\ideal,i}_{L_1G_1 K_* E}(\vec{\sigma}_{E|L_1 G_1}),
\rho^{\real,i}_{L_1G_1 K_* E})\nonumber \\
= &
\sum_{l} P_{L_1}(l)\max_{\vec{\sigma}_{E|G_1}}
F(\rho^{\ideal,i}_{G_1 K_* E|L_1=l}(\vec{\sigma}_{E| G_1}),
\rho^{\real,i}_{G_1 K_* E |L_1=l})\nonumber  \\
\stackrel{(a)}{\ge}  & \sum_{l} P_{L_1}(l) \sqrt{1- P_{du|L_1=l}^i}\nonumber  \\
\stackrel{(b)}{\ge}  &  \sqrt{\sum_{l} P_{L_1}(l) (1- P_{du|L_1=l}^i)} 
= \sqrt{1- P_{du}^i}\Label{XO},
\end{align}
where 
$(a)$ follows from \eqref{NM1}
and 
$(b)$ follows from the concavity of the function $x \mapsto \sqrt{x}$.
Thus, we have
\begin{align}
& 
\min_{\vec{\sigma}_{E|LG}}
\frac{1}{2}\|\rho^{\ideal,i}_{L G K E}(\vec{\sigma}_{E|L G})
-\rho^{\real,i}_{LG K E}\|_1 \nonumber \\
\stackrel{(a)}{\le}&
\min_{\vec{\sigma}_{E|L_1 G}}
\frac{1}{2}\|\rho^{\ideal,i}_{L_1G K_2 E}(\vec{\sigma}_{E|L_1 G})
-\rho^{\real,i}_{L_1G K_2 E}\|_1 \nonumber \\
\stackrel{(b)}{=} &
\min_{\vec{\sigma}_{E|L_1 G_1G_2}}
\frac{1}{2}\|\rho^{\ideal,i}_{L_1G_1 G_2 K_2 E}(\vec{\sigma}_{E|L_1 G_1G_2})-\rho^{\real,i}_{L_1G_1 G_2 K_2 E}\|_1 \nonumber \\
\le & 
\min_{\vec{\sigma}_{E|L_1 G_1}}
\frac{1}{2}\|\rho^{\ideal,i}_{L_1G_1 G_2 K_2 E}(\vec{\sigma}_{E|L_1 G_1})-\rho^{\real,i}_{L_1G_1 G_2 K_2 E}\|_1 \nonumber \\
\stackrel{(c)}{=}  & 
\min_{\vec{\sigma}_{E|L_1 G_1}}
\frac{1}{2}\|\rho^{\ideal,i}_{L_1G_1 K_1 K_2 E}(\vec{\sigma}_{E|L_1 G_1})-\rho^{\real,i}_{L_1G_1 K_1 K_2 E}\|_1 \nonumber \\
\stackrel{(d)}{\le}  & 
\min_{\vec{\sigma}_{E|L_1 G_1}}
\sqrt{1-F(\rho^{\ideal,i}_{L_1G_1 K_1 K_2 E}(\vec{\sigma}_{E|L_1 G_1}),
\rho^{\real,i}_{L_1G_1 K_1 K_2 E})^2} \nonumber \\
=  & 
\sqrt{1-
\max_{\vec{\sigma}_{E|L_1 G_1}}
F(\rho^{\ideal,i}_{L_1G_1 K_1 K_2 E}(\vec{\sigma}_{E|L_1 G_1}),
\rho^{\real,i}_{L_1G_1 K_1 K_2 E})^2} \nonumber \\
\stackrel{(e)}{\le} &
\sqrt{1-(1-P_{du}^i)}
=\sqrt{P_{du}^i},
\end{align}
where
$(a)$ follows from the fact that $K_2$ is a part of $K$,
$(b)$ follows from the relation $G=(G_1G_2)$,
$(c)$ follows from the one-to-one relation between
$(K_1,K_2)$ and $(G_2,K_2)$,
$(d)$ follows from the general inequality 
$\frac{1}{2}\|\rho-\sigma\|\le \sqrt{1- F(\rho,\sigma)^2}$ \cite[(6.106)]{Kyoritsu},
and 
$(e)$ follows from \eqref{XO}.
Hence, we obtain \eqref{NL2}.

\section{Proof of \eqref{NM1}}\Label{A2}
For simplicity, we show \eqref{NM1} only the case with $i=1$.
Since $L_1$ is fixed to $l$, we omit $L_1=l$ in the following discussion.
For $s,t \in \bF_2^l$,
we define operators $l$-qubit system as
 \begin{align}
W(s,t):=\Big(\sum_{x' \in \bF_2^l} |x'+s\rangle \langle x'| \Big)
\Big(\sum_{x \in \bF_2^l} (-1)^{t\cdot x}|x\rangle \langle x| \Big),
 \end{align}
where $t \cdot x:= \sum_{j=1}^l t_j x_j$.
Then, by using a distribution $P_{XZ}$ on $\bF_2^{2l}$,
a generalized Pauli channel $\Lambda[P_{XZ}]$ is written as
\begin{align}
\Lambda[P_{XZ}](\rho):=\sum_{(s,t)\in \bF_2^{2l}}
P_{XZ}(s,t) W(s,t)\rho W(s,t)^\dagger. 
\end{align}

As shown in \cite[Section V-B]{Ha07},
the noisy channel can be considered as a generalized Pauli channel
by considering the virtual application of discrete twirling.
Also, the virtual application of discrete twirling does not change the joint state 
on Alice and Bob.
Hence, we can consider that Alice and Bob made 
the virtual application of discrete twirling.
That is, We can consider that the obtained keys $K_*$ and $\hat{K}_*$
are obtained via quantum communication via 
a generalized Pauli channel.
In this case, as shown in \cite[Appendix B]{Ha07},
Eve's state $\rho_{E|K_*=k}$ with public information $G$
is given as
\begin{align}
\rho_{E|K_*=k}= \sum_{x \in \bF_2^l}P_X(x) 
|P_{XZ},k,x\rangle \langle P_{XZ},k,x|,
\end{align}
where
\begin{align}
|P_{XZ},y,x\rangle:=\sum_{z \in \bF_2^l} (-1)^{z\cdot y} \sqrt{P_{Z|X}(z|x)}
|x,z\rangle.
\end{align}
While the system $E$ is composed of $2l$ qubits,
the first $l$ qubits do not have off-diagonal elements.
When the first and second $l$ qubits in $E$ 
are written by $E_1$ and $E_2$,
$E_1$ can be considered as a classical system.

We have
\begin{align}
\rho_{K_* E}= \sum_{k \in \bF_2^l}\frac{1}{2^l}|k\rangle \langle k|
\otimes \rho_{E|K_*=k}.
\end{align}
Then,
\begin{widetext}
\begin{align}
&\max_{\sigma_{E}}
 F(\rho_{K_*  E}, \rho_{K^*}\otimes \sigma_{E})
\nonumber \\
 =&
\max_{\sigma_{E}}
F \Big( \sum_{k \in \bF_2^l}\frac{1}{2^l}|k\rangle \langle k|
\nonumber \otimes
\sum_{x \in \bF_2^l}P_X(x)
 |P_{XZ},k,x\rangle \langle P_{XZ},k,x|,
\sum_{k \in \bF_2^l}\frac{1}{2^l}|k\rangle \langle k|
\otimes
\sigma_{E_1 E_2}\Big) \nonumber \\
 =&
\max_{\sigma_{E_2|E_1=x}}
F \Big( \sum_{k \in \bF_2^l}\frac{1}{2^l}|k\rangle \langle k|
\otimes
\sum_{x \in \bF_2^l}P_X(x)
|P_{XZ},k,x\rangle \langle P_{XZ},k,x|,
 \sum_{k \in \bF_2^l}\frac{1}{2^l}|k\rangle \langle k|
\otimes
\sigma_{E_1 E_2}\Big) .
\end{align}
Since 
\begin{align}
&(I\otimes I\otimes W(0,t))
\sum_{k \in \bF_2^l}\frac{1}{2^l}|k\rangle \langle k|
\otimes
\sum_{x \in \bF_2^l}P_X(x)
|P_{XZ},k,x\rangle \langle P_{XZ},k,x|
(I\otimes I\otimes W(0,t))^\dagger \nonumber \\
=&
\sum_{k \in \bF_2^l}\frac{1}{2^l}|k\rangle \langle k|
\otimes
|P_{XZ},k,x\rangle \langle P_{XZ},k,x|
\end{align}
for $t \in \bF_2^l$,
the minimizer for $\sigma_{E_1 E_2}$ can be assumed to be invariant 
for $I\otimes W(0,t)$.
That is, $\sigma_{E_1 E_2}$ has the form
$\sum_{x,z \in \bF_2^l}Q_{XZ}(x,z) |x,z\rangle \langle x,z|$.
Hence, 
\begin{align}
& \max_{\sigma_{E_1 E_2}}
F \Big( \sum_{k \in \bF_2^l}\frac{1}{2^l}|k\rangle \langle k|
\otimes
\sum_{x \in \bF_2^l}P_X(x)
|P_{XZ},k,x\rangle \langle P_{XZ},k,x|,\sum_{k \in \bF_2^l}\frac{1}{2^l}|k\rangle \langle k|
\otimes
\sigma_{E_1E_2}\Big) \nonumber \\
=&\max_{\sigma_{E_1 E_2}}
\sum_{k \in \bF_2^l}\frac{1}{2^l}
F \Big( 
\sum_{x \in \bF_2^l}P_X(x)
|P_{XZ},k,x\rangle \langle P_{XZ},k,x|,
\sigma_{E_1E_2}\Big) \nonumber \\
=&
\max_{Q_{XZ}}
\sum_{k \in \bF_2^l}\frac{1}{2^l}
\sum_{x \in \bF_2^l}
\sqrt{P_X(x)Q_X(x)}
F\Big( |P_{XZ},k,x \rangle  \langle P_{XZ},k,x | ,
\sum_{z \in \bf_2^l} Q_{Z|X}(z|x)|x,z\rangle \langle x,z|
\Big) \nonumber \\
=&
\max_{Q_{XZ}}
\sum_{k \in \bF_2^l}\frac{1}{2^l}
\sum_{x \in \bF_2^l}
\sqrt{P_X(x)Q_X(x)
\Big \langle P_{XZ},k,x\Big| 
\sum_{z \in \bf_2^l} Q_{Z|X}(z|x)|x,z\rangle \langle x,z|
\Big|P_{XZ},k,x\Big\rangle}\nonumber  \\
=&\max_{Q_{XZ}}
\sum_{k \in \bF_2^l}\frac{1}{2^l}
\sum_{x \in \bF_2^l}
\sqrt{P_X(x)Q_X(x)
\sum_{z \in \bF_2^l}
P_{Z|X=x}(z) Q_{Z|X=x}(z) }\nonumber \\
=&\max_{Q_{X}}
\sum_{k \in \bF_2^l}\frac{1}{2^l}
\sum_{x \in \bF_2^l}
\sqrt{P_X(x)Q_X(x)\max_{z \in \bF_2^l} P_{Z|X=x}(z) }\nonumber \\
=&\max_{Q_{X}}
\sum_{x \in \bF_2^l}
\sqrt{P_X(x)Q_X(x)\max_{z \in \bF_2^l} P_{Z|X=x}(z) }
\stackrel{(a)}{=}  
\sqrt{\sum_{x \in \bF_2^l}
P_X(x) \max_{z \in \bF_2^l} P_{Z|X=x}(z) } \nonumber \\
\ge &
\sqrt{\max_{z \in \bF_2^l} P_{Z}(z) }
\ge \sqrt{1-P_{du}^1},
\end{align}
where
$(a)$ follows from
the following relation;
Let $\{\alpha_i\}$ be general non-negative real numbers.
We have the following minimization for probability distribution $q_i$;
\begin{align}
\max_{q_i} \sum_{i} \sqrt{ \alpha_i q_i }
=\sqrt{\sum_i \alpha_i},
\end{align}
where the maximum is attained when $q_i=\frac{\alpha_i}{\sum_{i'}\alpha_{i'}}$.
Therefore, we obtain \eqref{NM1}.

\section{Proof of \eqref{EA7}}\Label{A4}
Using the definition of $\delta(p,\epsilon,m_1,m_2)$ given in \eqref{XL1},
we calculate the length of the generated keys as follows.
\begin{align}
&\beta (1-r_1)n_1-m_{1}(n_1,p_2)-m_{3} 
+
\beta (1-r_2)n_2-m_{2}(n_2,p_1)-m_{3}\nonumber \\
=&\beta (1-r_1)n_1-(1-r_1)n_1( h(p_2+\delta(p_2,\epsilon,(1-r_1) n_1,r_2 n_2)))\nonumber \\
&+\beta (1-r_2)n_2-(1-r_2)n_2( h(p_1+\delta(p_1,\epsilon,(1-r_2) n_2,r_1 n_1)))
\nonumber \\
&- 2\log n \nonumber \\
=&\beta (1-r_1)n_1-(1-r_1)n_1
\Big( h(p_2)+h'(p_2)\delta(p_2,\epsilon,(1-r_1) n_1,r_2 n_2))
+o(\frac{1}{\sqrt{n}})\Big)
\nonumber \\
&+\beta (1-r_2)n_2-(1-r_2)n_2 \Big( h(p_1)+h'(p_1)\delta(p_1,\epsilon,(1-r_2) n_2,r_1 n_1))+o(\frac{1}{\sqrt{n}})\Big)
- 2\log n \nonumber \\
=&(1-r_1)( \beta -h(p_2))n_1 + (1-r_2)( \beta -h(p_1))n_2 \nonumber \\
&-(1-r_1) h'(p_2)\delta(p_2,\epsilon,(1-r_1) n_1,r_2 n_2))n_1
-(1-r_2) h'(p_1)\delta(p_1,\epsilon,(1-r_2) n_2,r_1 n_1))n_2
+o(\sqrt{n})
\nonumber \\
=&(1-r_1)( \beta -h(p_2))n_1 + (1-r_2)( \beta -h(p_1))n_2 \nonumber \\
&-B(p_2,\epsilon) \sqrt{\frac{((1-r_1) n_1+r_2n_2)(1-r_1) n_1}{r_2 n_2}} 
-B(p_1,\epsilon) \sqrt{\frac{((1-r_2) n_2+r_1n_1) (1-r_2) n_2}{r_1 n_1}} 
+o(\sqrt{n}).
\end{align}
Hence, we obtain \eqref{EA7}.
\end{widetext}

\end{document}